\documentclass[showpacs,prl,aps,twocolumn]{revtex4}
\usepackage{amsmath,amsfonts,bm,graphicx,hyperref}

\newcommand{\llangle}{\langle\!\langle}
\newcommand{\rrangle}{\rangle\!\rangle}

\begin{document}
\title{Distributions of Waiting Times of Dynamic Single-Electron Emitters}
\author{Mathias Albert}
\author{Christian Flindt}
\author{Markus B\"uttiker}
\affiliation{D\'epartement de Physique Th\'eorique, Universit\'e de Gen\`eve, CH-1211 Gen\`eve, Switzerland}
\date{\today}

\begin{abstract}
The distribution of waiting times between elementary tunneling events is of fundamental importance for understanding the stochastic charge transfer processes in nanoscale conductors. Here we investigate the waiting time distributions (WTDs) of periodically driven single-electron emitters and evaluate them for the specific example of a mesoscopic capacitor. We show that the WTDs provide a particularly informative characterization of periodically driven devices and we demonstrate how the WTDs allow us to reconstruct the full counting statistics (FCS) of charges that have been transferred after a large number of periods. We find that the WTDs are capable of describing short-time physics and correlations which are not accessible via the FCS alone.
\end{abstract}

\pacs {02.50.Ey, 72.70.+m, 73.23.Hk}


\maketitle

\textit{Introduction}.--- Investigating the electrical fluctuations in a nanoscale conductor is an attractive method to probe and characterize the physical mechanisms and correlations that determine a given quantum transport process \cite{Bla00}. In one approach, the stochastic number of transferred particles during a long time interval, the so-called full counting statistics (FCS), is studied \cite{Lev93}. FCS already has a significant history in the theory of mesoscopic physics, and recent measurements of current fluctuations in submicron devices have shown that FCS is no longer just an interesting theoretical concept \cite{Reu03}. It is also becoming an important experimental tool to examine interaction and coherence effects in nanoscale systems under out-of-equilibrium conditions.

The FCS, however, does not provide a complete picture of a charge transport process on all relevant time scales. Conventionally, FCS is defined in the long-time limit, where a large number of charges have passed through the conductor. Only the zero-frequency components of the current fluctuations can then be addressed and important short-time physics may be lost. Recently, systematic theories of \emph{finite-frequency} FCS have been developed in order to calculate frequency-dependent noise and higher-order cumulants \cite{Ema07}, but at this point these methods are still restricted to systems without any explicit time dependence. An alternative and particularly intuitive characterization of the charge transfer process is provided by the distribution of time delays between subsequent physical events, also known as the waiting time distribution (WTD).

\begin{figure}
   \begin{center}
     \includegraphics[width=0.42\textwidth]{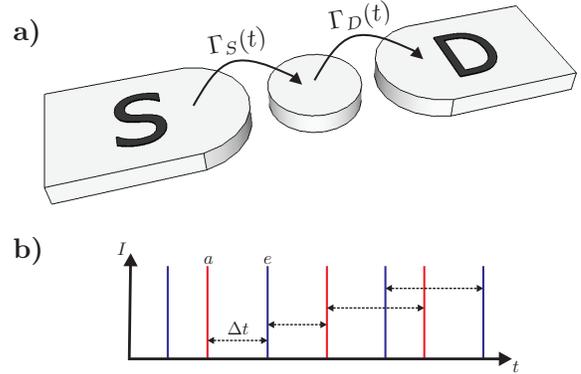}
     \caption{\label{fig_swt}(Color online) Single-electron emitter and waiting times. (a) Nanoscale system coupled to source ($S$) and drain ($D$) electrodes. Only a single electron at a time can occupy the nanoscale system, e.\ g., because of strong Coulomb interactions. Unidirectional transport takes place from source to drain due to the periodically modulated rates $\Gamma_S(t)$ and $\Gamma_D(t)$. (b) Current pulses in the source (blue [dark gray]) and drain (red [light gray]) electrodes together with illustrations of the corresponding waiting times $\Delta t$ (dashed lines) between elementary tunneling events, absorption ($a$) and emission ($e$).}
   \end{center}
 \end{figure}

While WTDs have been studied intensively in other fields of science, e.\ g., in single molecule chemistry \cite{Eng06,Gop06} and in quantum optics \cite{Car89,WtAtom}, they have only received very limited attention within the field of quantum transport. Exceptions include a few theoretical works on nondriven systems \cite{Sch05,Bra08,Wel08}, but a conceptually simple example of a nanoscale electronic system where the usefulness of WTDs is clearly demonstrated has to date been missing. In this Letter, we show that the WTDs of periodically driven single-electron emitters, Fig.\ \ref{fig_swt}, provide a very useful view on the charge transfer statistics and correlations in such systems. In particular, we evaluate the WTDs of a mesoscopic capacitor \cite{But93,Gab06,Mah10}, which serves as a prime example of the usefulness of WTDs. We derive expressions for the WTDs which are applicable also to certain types of quantum pumps \cite{Blu07,Sin07,Bat10} and nanoelectromechanical systems \cite{Pis04}.  We demonstrate how the WTDs not only allow us to reconstruct and obtain the FCS of emitted electrons, but additionally they contain information about the charge transfer process on short time-scales which is not available in the FCS alone. As we show the WTDs describe the charge transfer process on \emph{all} important time scales.

\textit{System}.---  We focus on systems consisting of a source and a drain electrode coupled to a nanoscale conductor, Fig.\ \ref{fig_swt}(a), biased such that single-electron transport is unidirectional from the source to the drain.  The tunneling rates to and from the conductor, $\Gamma_{S}(t)$ and $\Gamma_{D}(t)$, respectively, are time dependent. The probability $P(t)$ for the conductor to be occupied by an electron obeys the evolution equation
\begin{equation}\label{eq_model1}
   \partial_t P(t)  = -\Gamma_D(t)P(t)+\Gamma_S(t)[1-P(t)]\, ,
 \end{equation}
where $1-P(t)$ is the probability for the conductor to be empty. This model suffices to illustrate the basic concepts of WTDs which are of interest here. We concentrate on systems, similar to the ones mentioned in the Introduction, where the tunneling rates are periodic in time, such that $\Gamma_\alpha(t)=\Gamma_{\alpha}(t+T)$, $\alpha=S,D$, with $T$ being the period.

\textit{Waiting time distributions}.--- We consider the waiting times between different tunneling events, Fig.\ \ref{fig_swt}(b). These consist of events, where the conductor either absorbs an electron from the source or emits an electron via the drain. Because of the probabilistic nature of the charge transfer, the waiting time $\Delta t$ between such events is itself a random variable. We use  $w_{ea}(\Delta t,t_0)$ [$w_{ae}(\Delta t,t_0)$] to denote the probability for the first emission event to occur at time $\Delta t+t_0$ given that absorption occurred at the random earlier time $t_0$ [and similar for absorption following emission]. The same definitions apply to the WTDs for tunneling events of the same type, $w_{ee}(\Delta t, t_0)$ and $w_{aa}(\Delta t, t_0)$. Since the occupation of the conductor is either 0 or 1, two events of the same kind cannot happen simultaneously and $w_{ee}(0, t_0)=w_{aa}(0, t_0)=0$ for all $t_0$. For nondriven systems, where the tunneling rates are time-independent, translational invariance with respect to time implies that the WTDs do not depend on $t_0$ \cite{Bra08}.

To calculate the WTDs we first express the source and drain mean currents at time $t_0$ as $\langle I_{S}(t_0)\rangle=\Gamma_S(t_0)[1-P(t_0)]$ and $\langle I_{D}(t_0)\rangle=\Gamma_D(t_0) P(t_0)$. The currents are proportional to the probabilities of absorbing and emitting an electron, respectively. Additionally, we need the \emph{conditional} currents, e.\ g., $\langle I^a_{D}(\Delta t,t_0)\rangle=\Gamma_D(t_0+\Delta t)P^a(\Delta t,t_0)$. Here, $P^a(\Delta t,t_0)$ is the survival probability of an electron at time $t_0+\Delta t$ given that it was absorbed at time $t_0$, such that $P^a(0,t_0)=1$. The survival probability $P^a(\Delta t,t_0)$ obeys Eq.\ (\ref{eq_model1}) with $\Gamma_S(t)=0$, i.\ e., $\partial_{\Delta t}P^a(\Delta t,t_0)=-\Gamma_D(t_0+\Delta t)P^a(\Delta t,t_0)$. The WTD between absorption and emission is now $w_{ea}(\Delta t,t_0)=\mathcal N_{ea} \langle I_{S}(t_0)\rangle \langle I^a_{D}(\Delta t,t_0)\rangle$, where $\mathcal N_{ea}$ is a normalization constant. The WTD is $T$-periodic in the second time argument, $w_{ea}(\Delta t,t_0+T)=w_{ea}(\Delta t,t_0)$. This allows us to consider only the finite time interval $t_0\in [0,T]$ and choose the normalization constant $\mathcal N_{ea}$, such that $\int_{0}^{T} dt_0\int_0^{\infty} d(\Delta t)w_{ea}(\Delta t,t_0)=1$.

At this point, the WTD depends not only on the waiting time $\Delta t$, but also on the absolute time $t_0$ at which absorption occurred. We are only interested in the waiting time itself and therefore integrate out $t_0$. Defining $\overline{f}=\int_0^T dt_0 f(t_0)$ for a $T$-periodic function $f(t_0)$, we find for the WTD between absorption and emission
\begin{equation}
\overline{w}_{ea}(\Delta t)=\mathcal N_{ea}\overline{\langle I_{S}(t_0)\rangle \langle I^a_{D}(\Delta t,t_0)\rangle}.
\end{equation}
Similar reasoning for the WTD between emission and absorption leads to the equivalent expression
\begin{equation}
\overline{w}_{ae}(\Delta t)=\mathcal N_{ae}\overline{\langle I_{D}(t_0)\rangle \langle I^e_{S}(\Delta t,t_0)\rangle},
\end{equation}
with the conditional current $\langle I^e_{S}(\Delta t,t_0)\rangle=\Gamma_S(t_0+\Delta t)P^e(\Delta t,t_0)$. Here $P^e(\Delta t,t_0)$ is the survival probability of the empty state. Proceeding along the same lines for the WTDs between events of the same kind, we find
\begin{widetext}
\begin{equation}
\begin{split}
   \overline{w}_{aa}(\Delta t)= &\mathcal N_{aa} \int_0^{\Delta t}d(\Delta t') \overline{\langle I_{S}(t_0)\rangle \langle I^a_{D}(\Delta t',t_0)\rangle\langle I^e_{S}(\Delta t-\Delta t',t_0+\Delta t')\rangle},\\
   \overline{w}_{ee}(\Delta t)= &\mathcal N_{ee} \int_0^{\Delta t}d(\Delta t') \overline{\langle I_{D}(t_0)\rangle \langle I^e_{S}(\Delta t',t_0)\rangle \langle I^a_{D}(\Delta t-\Delta t',t_0+\Delta t')\rangle},
\end{split}
\end{equation}
\end{widetext}
where the constants $\mathcal N_{aa}$ and $\mathcal N_{ee}$ ensure the normalizations $\int_0^\infty d(\Delta t)\overline{w}_{aa}(\Delta t) = 1$ and $\int_0^\infty d(\Delta t)\overline{w}_{ee}(\Delta t) = 1$, respectively. In the expression for $\overline{w}_{aa}(\Delta t)$ [$\overline{w}_{ee}(\Delta t)$], $\Delta t'$ is the time interval between the first absorption [emission] event and the intermediate emission [absorption] event which finally is followed by the second absorption [emission] event after the total waiting time $\Delta t$.

\begin{figure*}
    \begin{center}
      \includegraphics[width=0.99\linewidth]{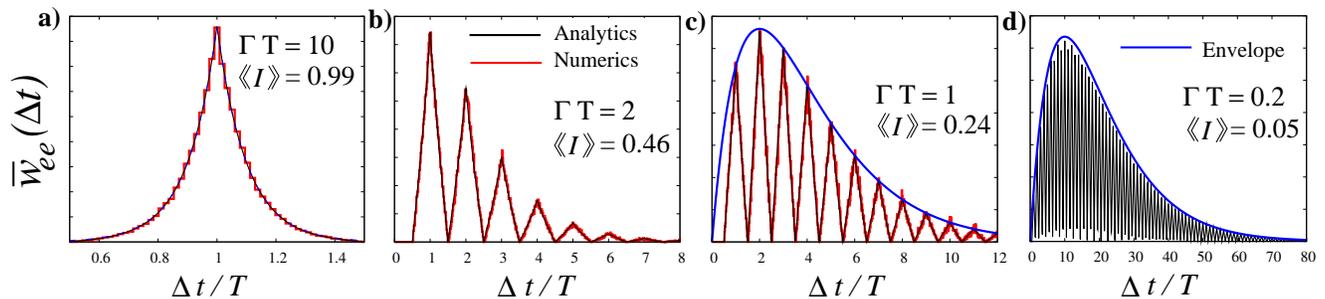}
      \caption{\label{fig_wt_qc} (Color online). Waiting time distribution (WTD) for the mesoscopic capacitor. We show the WTD between subsequent emission events $\overline{w}_{ee}(\Delta t)$ for different values of the tunneling rate $\Gamma$ in units of the inverse period of the driving $T^{-1}$. The mean charge emitted per period (the mean current) $\llangle I\rrangle$ is also indicated. The analytic result, given by Eq.\ (\ref{eq_wt_qc3}), is compared to numerical simulations of the charge transport. For large values of $\Gamma$, (a), the charge transport is highly regular and periodic with the mean waiting time equal to the period, $\llangle\Delta t\rrangle=T$. As $\Gamma$ is reduced,  (b), several peaks appear in the WTD. For even smaller values of $\Gamma$, (c) and (d), the mean waiting time is much larger than the period, $\llangle\Delta t\rrangle\gg T$, and the overall shape of the WTD is determined by the envelope curve shown in blue. The mean charge emitted per period is then much smaller than 1.}
    \end{center}
\end{figure*}

\textit{Mesoscopic capacitor}.---  We now illustrate our findings in terms of a specific example: a mesoscopic capacitor consisting of a nanoscale cavity coupled to external reservoirs via a quantum point contact \cite{But93,Gab06}. When  subject to fast periodic gate voltage modulations, the capacitor can absorb and re-emit single electrons at giga-hertz frequencies. The system can be described by Eq.\ (\ref{eq_model1}) taking $\Gamma_S(t)=\Gamma$ and $\Gamma_D(t)=0$ in the first half of the period, and $\Gamma_S(t)=0$ and $\Gamma_D(t)=\Gamma$ in the second half \cite{Mah10,Alb10}. The tunneling rate $\Gamma$ can be controlled experimentally by adjusting electrostatically the opening of the quantum point contact.

Following the steps described above we obtain a simple expression for the WTD between emission and absorption events
  \begin{equation}\label{eq_wt_qc1}
    \overline w_{ae}(\Delta t)=\frac{\Gamma\, \varepsilon^{\lfloor \Delta t/T\rfloor}}{2(1-\varepsilon)}\left\{e^{-|\xi_{ae}|}-\varepsilon^2 e^{|\xi_{ae}|}\right\},
  \end{equation}
where $\varepsilon=e^{-\Gamma\, T/2}$, $\xi_{ae}=\Gamma [\Delta t-(\lfloor \Delta t/T\rfloor+1/2)T]$, and $\lfloor x\rfloor$ is the integer part of $x$. The result contains two independent structures: an internal structure (in curly brackets) which is periodic with $T$ and an envelope (given by $\varepsilon^{\lfloor \Delta t/T\rfloor}$) which is responsible for an exponential decay of the WTD for large waiting times $\Delta t$. For the WTD between two emission events we find
  \begin{equation}\label{eq_wt_qc3}
    \overline w_{ee}(\Delta t)=\frac{\Gamma\lfloor\frac{\Delta t+T/2}{T}\rfloor \varepsilon^{\lfloor\frac{\Delta t-T/2}{T}\rfloor }}{2}\left\{e^{-|\xi_{ee}|}-\varepsilon^2 e^{|\xi_{ee}|}\right\}
  \end{equation}
with $\xi_{ee}=\Gamma [\Delta t-(\lfloor \Delta t/T\rfloor+1)T]$. Again, the WTD consists of an envelope function and an internal structure.

Our analytic results are confirmed by numerical simulations of the mesoscopic capacitor, Fig.\ \ref{fig_wt_qc}. For large tunneling rates,  Fig.\ \ref{fig_wt_qc}(a), the transport process is predominantly regular and periodic with one electron emitted almost every cycle. The WTD has a single peak centered around the period $\Delta t \simeq T$. The peak, however, is not sharp due to the jitter in the emission process, causing phase noise \cite{Mah10}. As the tunneling rate is decreased, Fig.\ \ref{fig_wt_qc}(b), a more complicated structure appears with several equidistant peaks separated by the period. Two emission events must be separated by at least half a period implying that $\overline w_{ee}(\Delta t)=0$ for $\Delta t<T/2$. Interestingly, this short-time behavior is not visible in the noise spectrum of the capacitor found in Ref.\ \cite{Alb10}. For even smaller tunneling rates, Figs.\ \ref{fig_wt_qc}(c) and \ref{fig_wt_qc}(d), the charge transport becomes increasingly random, and subsequent electron emissions are typically separated by several periods. The current fluctuations are then shot-noise like and the overall shape of the WTD is determined by the envelope function of the approximate form $\Delta t e^{-(\Gamma/2) \Delta t}$. This corresponds to the WTD for the case, where both tunneling rates are kept constant as $\Gamma_S(t)=\Gamma_D(t)=\Gamma/2$ for all $t$.  As we show below, much of this information is not available in the FCS alone.

\textit{Full counting statistics}.---  To connect the WTDs to the FCS, the probability $\mathcal P(n,N)$ of emitting $n$ electrons during a large number of periods $N$, we assume that maximally a single electron can be emitted during a period. This is the case for the mesoscopic capacitor considered above. We can then write the probability distribution as $\mathcal P(n,N)=\sum_{m_1,\ldots,m_n}\widetilde{w}_{ee}(m_1)\cdots\widetilde{w}_{ee}(m_n)\delta_{m_1+\ldots+m_n,N}$,
where $\widetilde{w}_{ee}(m)$ is the \emph{coarse-grained} WTD for the number of periods $m$ between two subsequent emission events. The Kronecker delta $\delta_{m,N}$ expresses the constraint that the sum of periods between emission events $m_1+\ldots+m_n$ must equal the total number of periods $N$. We have assumed that the counting of emitted electrons starts right after an emission event, but the specific choice of initial condition is not important for the FCS after a large number of periods.

Next, we introduce the cumulant generating function (CGF) $\mathcal S(\chi,N)=\log\sum_{n} \mathcal P(n,N) e^{i\chi n}$ whose derivatives with respect to the counting field $\chi$ at $\chi=0$ yield the cumulants of $n$ as $\llangle n^k\rrangle =\partial^k_{i\chi} \mathcal S(\chi,N)|_{\chi\rightarrow 0}$. Additionally, we define the discrete Laplace transform of $\widetilde{w}_{ee}(m)$, $\hat{\widetilde{w}}_{ee}(z)=\sum_m \widetilde{w}_{ee}(m) e^{-mz}$ and the corresponding CGF of $m$, $\mathcal W_{ee} (z)=\log \hat{\widetilde{w}}_{ee}(-z)$, which similarly delivers the cumulants of $m$ as $\llangle m^k\rrangle=\partial^k_z \mathcal W_{ee} (z)|_{z\rightarrow 0}$.
We can then express the CGF of $n$ as
\begin{equation}\label{eq_wt_fcs3}
  e^{\mathcal S(\chi,N)}=\frac{1}{2\pi i}\int_{-i\pi }^{i\pi}dz \frac{e^{z  N}}{1-e^{i\chi+\mathcal W_{ee} (-z)}}\,.
\end{equation}
In the large-$N$ limit, the integral is determined by the particular pole $z=z_0(\chi)$ with $z_0(0)=0$ that solves the equation
\begin{equation}\label{eq_wt_fcs4}
 i\chi+\mathcal W_{ee}(-z)=0\,,
\end{equation}
such that $\mathcal S(\chi,N)\rightarrow N z_0(\chi)$. The electron current $I\equiv n/N$ is the number of emitted electrons $n$ over the number of periods $N$, and $z_0(\chi)$ thus generates the cumulants of the current, i.\ e., $\llangle I^k\rrangle=\partial^k_{i\chi} z_0(\chi)|_{\chi\rightarrow 0}$. For the mesoscopic capacitor, the large-$N$ limit is reached for $N\gg \max\{1,1/\Gamma T\}$.

Equation (\ref{eq_wt_fcs4}) demonstrates an important and intimate connection between fluctuations in the current of emitted electrons $I$ and the number of periods $m$  between  emission events. In general, the equation may be difficult to solve for $z=z_0(\chi)$, but it provides us with a simple and systematic way of relating the current cumulants to the cumulants of $m$: taking consecutive derivatives of the left hand side with respect to the counting field $\chi$ evaluated at $\chi=0$, we find for the average current $\llangle I\rrangle = 1/\llangle m\rrangle$ and the (normalized) current cumulants $F_k=\llangle I^k\rrangle/\llangle I\rrangle$
\begin{equation}
\label{eq_wt_fcs6}
   \begin{split}
   F_2& =\frac{\llangle m^2\rrangle}{\llangle m\rrangle^2},\\
   F_3&=3\frac{\llangle m^2\rrangle^2}{\llangle m\rrangle^4}-\frac{\llangle m^3\rrangle}{\llangle m\rrangle^3},\\
   F_4&= 15 \frac{\llangle m^2\rrangle^3}{\llangle m\rrangle^6} - 10 \frac{\llangle m^2\rrangle \llangle m^3\rrangle}{\llangle m\rrangle^5}+\frac{\llangle m^4\rrangle}{\llangle m\rrangle^4}.
   \end{split}
 \end{equation}
For non-driven systems, the number of periods  $m$ should be replaced by the continuous waiting time $\Delta t$. We then recover the known relations $\llangle I\rrangle =1/\llangle \Delta t\rrangle$ and $F_2  =\llangle \Delta t^2\rrangle/\llangle \Delta t\rrangle^2$ \cite{WtAtom}, see also Ref.\ \cite{Gop06}. Our derivation, however, allows us to determine current cumulants of \emph{any} order.

\begin{figure}
    \begin{center}
      \includegraphics[width=0.42\textwidth]{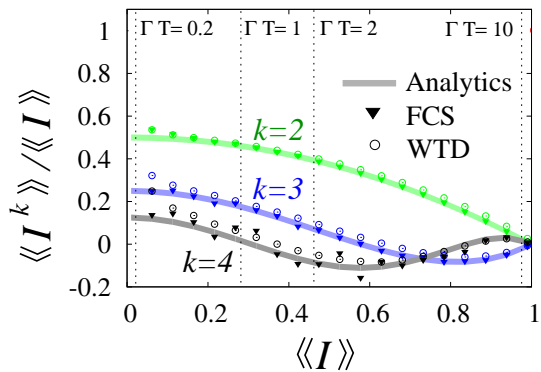}
      \caption{\label{fig_fcs} (Color online). FCS of the mesoscopic capacitor. The normalized current cumulants $F_k=\llangle I^k\rrangle/\llangle I\rrangle$, $k=2,3,4$, were obtained from numerical simulations (FCS) as well as from Eq.\ (\ref{eq_wt_fcs6}) using both numerical (WTD) and analytical (Analytics) results for the coarse-grained WTD. Dashed lines indicate values of $\Gamma T$ corresponding to the panels of Fig.\ \ref{fig_wt_qc}.}
    \end{center}
\end{figure}

In Fig.\ \ref{fig_fcs} we show the FCS for the mesoscopic capacitor. We performed separate numerical simulations of the FCS and the WTD, and for comparison we then used Eq.\ (\ref{eq_wt_fcs6}) to obtain the normalized current cumulants $F_k$ from the coarse-grained WTD.  Additionally, from Eq.\ (\ref{eq_wt_qc3}) we found analytically the CGF of $m$, $\mathcal W_{ee}(z)=z+2\log[(1-\varepsilon)/(1-\varepsilon e^z)]$, and again used Eq.\ (\ref{eq_wt_fcs6}) to obtain the $F_k$'s. The figure confirms the validity of Eq.\ (\ref{eq_wt_fcs6}) and clearly illustrates that the FCS can be obtained from the coarse-grained WTD. Importantly, the procedure cannot be reversed: The WTD cannot be obtained from the FCS. Moreover, comparing Figs.\  \ref{fig_wt_qc} and \ref{fig_fcs}, substantial information about the charge transfer process is obviously lost in the FCS. In the phase noise regime, $\Gamma T=10$ [corresponding to Fig.\ \ref{fig_wt_qc}(a)], the current cumulants are close to zero due to the regular emission of electrons. However, contrary to the WTD, the cumulants are not sensitive to the jitter in the emission process. As the tunneling rate is lowered, $\Gamma T=2$ [Fig.\ \ref{fig_wt_qc}(b)], several peaks appear in the WTD, but this is also not visible in the FCS, neither is the fact that two emission events must be separated by at least half a period. In the shot-noise regime, $\Gamma T=1,0.2$ [Figs.\ \ref{fig_wt_qc}(c) and \ref{fig_wt_qc}(d)], the cumulants approach the limiting values $F_k\rightarrow (1/2)^{k-1}$ corresponding to a Poisson process with an effective charge of $1/2$. This is also a very different characterization compared to the one provided by the WTDs.

\textit{Conclusions}.--- We have shown that the distribution of waiting times between elementary tunneling events is a useful tool to probe and characterize the charge fluctuations and correlations of periodically driven single-electron emitters on all important time scales. As a specific example, we have considered a mesoscopic capacitor for which we demonstrated that the WTDs contain considerable additional information compared to what is available in the FCS alone.

\textit{Acknowledgments}.--- We thank E.\ Bocquillon, T.\ Brandes, J.\ Dalibard, P.\ Degiovanni, G.\ F\`eve,  G.\ Haack, D.\ Kambly, M.\ Moskalets, and F.\ D.\ Parmentier for useful discussions. The work was supported by Swiss NSF and the NCCR Quantum Science and Technology.


\begin{thebibliography}{99}
\bibitem{Bla00} Ya. Blanter and M. B\"{u}ttiker, Phys. Rep. {\bf 336}, 1 (2000).
\bibitem{Lev93} L. S. Levitov and G. B. Lesovik, JETP Lett. {\bf 58}, 230 (1993), Yu. V. Nazarov (ed.), \emph{Quantum Noise in Mesoscopic
Physics} (NATO Science Series, Kluwer, Dordrecht, 2003).
\bibitem{Reu03} B. Reulet, J. Senzier, and D. E. Prober, Phys. Rev. Lett. {\bf 91}, 196601 (2003), Yu. Bomze {\em et al.}, Phys. Rev. Lett. {\bf 95}, 176601 (2005), S. Gustavsson {\em et al.}, Phys. Rev. Lett. {\bf 96}, 076605 (2006), T. Fujisawa {\em et al.}, Science {\bf 312}, 1634 (2006), E. V. Sukhorukov {\em et al.}, Nat.\ Phys. {\bf 3}, 243 (2007), C. Flindt {\em et al.}, Proc. Natl. Acad. Sci. USA {\bf 106}, 10116 (2009), J. Gabelli and B. Reulet, Phys. Rev. B {\bf 80}, 161203(R) (2009).
\bibitem{Ema07} C. Emary {\em et al.}, Phys. Rev. B {\bf 76}, 161404(R) (2007),  D. Marcos {\em et al.}, New J. Phys. {\bf 12}, 123009 (2010).
\bibitem{Eng06} B. P. English {\em et al.}, Nat. Chem.\ Biol.\ {\bf 2}, 87 (2006).
\bibitem{Gop06}  I. V. Gopich and A. Szabo, J. Chem. Phys. {\bf 124}, 154712 (2005).
\bibitem{Car89} H.\ J.\ Carmichael {\em et al.}, Phys. Rev. A {\bf 39}, 1200 (1989).
\bibitem{WtAtom} C. Cohen-Tannoudji and J. Dalibard, Europhys. Lett. \textbf{1}, 441 (1986), S. Reynaud, J. Dalibard and C. Cohen-Tannoudji, IEEE J. Quantum Electron. \textbf{24}, 1395 (1988).
\bibitem{Sch05} J. Schriefl {\em et al.}, Phys. Rev. B {\bf 72}, 035328 (2005).
\bibitem{Bra08} T. Brandes, Ann. Phys. (Berlin) {\bf 17}, 477 (2008).
\bibitem{Wel08} S. Welack {\em et al.}, Phys. Rev. B {\bf 77}, 195315 (2008), Europhys. Lett. \textbf{85}, 57008 (2009).
\bibitem{But93} M. B\"uttiker, H. Thomas, and A. Pr\^etre, Phys. Lett. A {\bf 180}, 364 (1993).
\bibitem{Gab06} J. Gabelli {\em et al.}, Science {\bf 313}, 499 (2006), G. F\`eve {\em et al.}, Science {\bf 316}, 1169 (2007).
\bibitem{Mah10} A. Mah\'e {\em et al.}, Phys. Rev. B {\bf 82}, 201309(R) (2010).
\bibitem{Blu07} D. Blumenthal \emph{et al.}, Nat. Phys. {\bf 3}, 343 (2007), N. Maire \textit{et al.}, Appl. Phys. Lett. \textbf{92}, 082112 (2008).
\bibitem{Sin07} N. A. Sinitsyn and I. Nemenman, Phys. Rev. Lett. {\bf 99}, 220408 (2007).
\bibitem{Bat10} F. Battista and P. Samuelsson, Phys. Rev. B {\bf 83}, 125324 (2011).
\bibitem{Pis04} F. Pistolesi,  Phys. Rev. B {\bf 69}, 245409 (2004).
\bibitem{Alb10} M. Albert, C. Flindt and M. B\"uttiker, Phys. Rev. B {\bf 82}, 041407(R) (2010).

\end{thebibliography}
\end{document}